\documentclass[a4paper, amsfonts, amssymb, amsmath, reprint, showkeys, nofootinbib, twoside]{revtex4-1}
\usepackage[english]{babel}
\usepackage[utf8]{inputenc}
\usepackage[colorinlistoftodos, color=green!40, prependcaption]{todonotes}
\usepackage{amsthm, bm}
\usepackage{mathtools}
\usepackage{physics}
\usepackage{xcolor}
\usepackage{graphicx}
\usepackage[left=18mm,right=18mm,top=35mm,columnsep=15pt]{geometry} 
\usepackage{adjustbox}
\usepackage{placeins}
\usepackage[T1]{fontenc}
\usepackage{lipsum}
\usepackage{csquotes}
\usepackage[pdftex, pdftitle={Article}, pdfauthor={Author}]{hyperref} 
\begin{document}
\title{Drag force near the QCD critical point}

\author{Yukinao Akamatsu}
    \email[Correspondence email address: ]{yukinao.a.phys@gmail.com}
\author{Masayuki Asakawa}
    \affiliation{Department of Physics, Osaka University, Toyonaka, Osaka 560-0043, Japan}

\date{\today}

\begin{abstract}
We discuss how heavy quark dynamics is affected by the critical fluctuations near the QCD critical point at finite temperature and density.
We find that the heavy quark momentum diffusion constant scales as $\kappa\propto\xi^{z-3-\eta}$.
In the model H scenario, which is widely accepted for the critical dynamics, the exponents are known as $z\simeq 3, \eta\sim 0.04$ and the critical singularity of $\kappa$ is not significant if any.
In the model B scenario, $z\simeq 4$ and $\kappa\propto \xi$ is singular near the critical point.
\end{abstract}


\maketitle

\section{Introduction} \label{sec:intro}
Now that the existence of the quark-gluon plasma (QGP) has been experimentally confirmed \cite{Arslandok:2023utm}, the next step is to determine the parameters in the phase diagram of QCD. 
Among them of particular interest is the position of the critical point if it exists \cite{Asakawa:1989bq}. 
To confirm the existence of the critical point and determine its position on the phase diagram, BES and BES II experiments have been carried out at RHIC and several experiments are planned at other facilities.
Several observable have been proposed for the critical point search.
One of them is characteristic behavior of higher cumulants of net baryon charge as a function of the energy of colliding nuclei.
However, this expectation is based on several unphysical assumptions such as freezeout of baryon number cumulants at chemical freezeout and no diffusion effect on them in the following hadron phase \cite{Asakawa:2019kek}.
So far, no definite observables exist as signatures of the critical point.
Thus, it would be nice to examine the possibility to use other observables for this purpose. 

In this paper, we study possibility to use a dynamical critical phenomenon on heavy quarks (in the following, we understand that heavy quark means (anti)charm quark, since (anti)bottom quarks are rarely produced in low energy heavy ion collisions), increase of the momentum diffusion constant around the critical point.  
The elliptic flow, $v_2$, is monotonously increasing in time.
If the momentum diffusion constant diverges around the critical point as the drag force on particles does in many models,  the motion of the heavy quarks would be synchronized with the motion of the bulk matter when the system passes near the critical point.
As a result, the $v_2$ of heavy quarks will show a peak structure at the collision energy at which system passes the critical point, as the collision energy is changed. 

This critical phenomenon is interesting also from the theoretical point of view.
The pursued critical point is for the chiral symmetry.
It is believed to belong to the dynamical universality class H in the classification by Hohenberg and Halperin \cite{RevModPhys.49.435, Son:2004iv}. 
Heavy quarks interact with gluons in completely the same manner as light quarks, but the concept of the chiral symmetry is not applied to heavy quarks.
In other words, heavy quarks are outside of the chiral symmetry. 
Thus heavy quarks acts as impurity when the bulk system goes through the chiral phase transition.
However, it obeys the same interaction law with light quarks that constitute the bulk.
This is the unique feature of heavy quarks as impurity.
In most cases, impurity interacts differently from the constituents of the matter that is going through a critical phenomenon. 
At this point it would be useful to imagine and compare with pollens in Brown's experiment \cite{brown1828xxvii}. 

In the following, we first model the coupling between heavy quarks and soft mode.
Then, we discuss the critical behavior of the momentum diffusion constant around the critical point in model H and model B.
Finally, we give a conclusion and an outlook.

\section{Coupling between heavy quarks and soft mode} \label{sec:coupling}

It is not well understood how heavy quarks couple to critical fluctuations near the QCD critical point.
In this paper, we construct the interaction Lagrangian $\mathcal L_\text{I}$ based on symmetry principles of QCD.
From the rotational invariance, we consider the heavy quark coupling in the (pseudo)scalar channels $\bar\psi\psi, \bar\psi\gamma_0\psi, \bar\psi\gamma_5\psi, \bar\psi\gamma_5\gamma_0\psi$.
Of these, the last two channels vanish in the heavy quark mass limit and the non-vanishing scalar channels are
\begin{align}
\bar\psi\psi &= Q^{\dagger} Q + Q_c^{\dagger}Q_c, \\
\bar\psi\gamma_0\psi &= Q^{\dagger} Q - Q_c^{\dagger}Q_c,
\end{align}
where $\psi = (Q, Q_c^*)^T$ in the Dirac representation.
Note that the vector channels inevitably couple to the critical mode through derivative interaction, so it is expected to be less important in the heavy quark momentum diffusion.  

When up and down quarks are massless and strange quark is somewhat heavy and does not play an essential role in the phase diagram, the critical fluctuations are $\vec\phi =(\sigma, \pi_a) \ (a=1,2,3)$ and form $\text{SU(2)}_\text{L}\times \text{SU(2)}_\text{R}\simeq \text{O}(4)$ quartet \cite{Rajagopal:1992qz}.
A small but finite mass $m$ for up and down quarks breaks the $\text{SU(2)}_\text{L}\times \text{SU(2)}_\text{R}$ symmetry explicitly, which makes $\pi_a$ massive near the critical point.
Note that at the critical point for finite $m$, $\sigma$ mixes with conserved densities of baryon number and energy and forms a soft mode.
This mixing is crucial to determine the dynamical universality to be model H (critical fluid) \`a la Hohenberg and Halperin \cite{RevModPhys.49.435, Son:2004iv}, where this soft mode couples to transverse momentum density nonlinearly.

Here, we construct the interaction Lagrangian for heavy quarks when up and down quarks are massive, but we can repeat the similar argument to discuss when they are both massless.
We use $\text{O}(4)$ symmetry and its small explicit breaking by finite $m$ as a guiding principle to write down the interaction Lagrangian.
We only consider the coupling between $\vec\phi$ and heavy quarks.
This is enough because $\sigma$ contains the soft mode.
Near the critical point, the effective interaction Lagrangian takes the form of
\begin{align}
\mathcal L_\text{I} &= -f(\vec \phi; T, \mu, m) \bar\psi\psi - g(\vec \phi; T, \mu, m) \bar\psi\gamma_0\psi,
\label{eq:interaction}
\end{align}
where ${\rm O}(4)$ symmetry constrains
\begin{align}
f(\vec \phi; T, \mu, m) &=f_0(|\vec \phi|^2; T, \mu) + \mathcal O(m), \label{eq:coeff_f} \\
g(\vec \phi; T, \mu, m) &=g_0(|\vec \phi|^2; T, \mu) + \mathcal O(m). \label{eq:coeff_g}
\end{align}
Charge conjugation symmetry also relates
\begin{align}
f(\vec \phi; T, \mu, m) &= f(\vec \phi^C; T, -\mu, m), \\
g(\vec \phi; T, \mu, m) &= -g(\vec \phi^C; T, -\mu, m),
\label{eq:C_g}
\end{align}
where $\vec \phi^C$ is the charge conjugation of $\vec \phi$.
At vanishing chemical potential, Eq.~\eqref{eq:C_g} constrains $g_0(|\vec \phi|^2; T, \mu=0)=0$ using $|\vec \phi^C|^2 = |\vec \phi|^2$.
However, at finite $\mu$, the charge conjugation does not constrain the interaction Lagrangian \eqref{eq:interaction}.

Since $\pi_a$ becomes massive due to small explicit breaking by finite $m$, we can concentrate only on the critical mode $\delta\sigma \equiv \sigma - \sigma_c$ to get
\begin{align}
f_0(|\vec \phi|^2; T_c, \mu_c) &\simeq f_0((\sigma_c + \delta\sigma)^2; T_c, \mu_c), \\
g_0(|\vec \phi|^2; T_c, \mu_c) &\simeq g_0((\sigma_c + \delta\sigma)^2; T_c, \mu_c).
\end{align}
We assume that the functions $f_0, g_0$ are regular and can be approximated by
\begin{align}
f_0((\sigma_c + \delta\sigma)^2; T_c, \mu_c) &\simeq f_0(\sigma_c^2) + 2f_0'(\sigma_c^2) \sigma_c \delta\sigma,\\
g_0((\sigma_c + \delta\sigma)^2; T_c, \mu_c) &\simeq g_0(\sigma_c^2) + 2g_0'(\sigma_c^2) \sigma_c \delta\sigma.
\end{align}
Mean field analysis around a tricritical point shows $\sigma_c\propto m^{0.2}$ when $m$ is small \cite{Fujii:2004jt}.
Thus the interaction Lagrangian is
\begin{align}
\mathcal L_\text{I} &=  -\delta\sigma (\lambda  Q^{\dagger}Q + \lambda_c Q_c^{\dagger}Q_c)  + \mathcal O(m,\delta\sigma^2),
\end{align}
where higher order fluctuations are less singular, e.g. $\delta\sigma^2$ is local composite field corresponding to energy fluctuations in the Ising model.

The coupling between heavy quark density and $\delta\sigma$ occurs by the same reason that baryon number density and $\delta\sigma$ mixes at finite chemical potential.
It can also be understood from the hadronic interaction at finite density.
In the vacuum, the heavy quark current couples to omega meson by $\omega_{\mu}\bar\psi\gamma^{\mu}\psi$.
There arises a mixing of (temporal component of) omega and sigma mesons at finite density \cite{CHIN1977301}, which induces the coupling between heavy quark density and sigma meson as we obtain above.

When up and down quarks are massless $m=0$, the condensate vanishes at the critical point $\sigma_c=0$ and all four components of $\vec\phi$ are critical.
In this case, the $\text{O}(4)$ symmetry constrains the leading terms in $\mathcal L_\text{I}$ to be $\propto |\vec\phi|^2$ (see Eqs.~\eqref{eq:coeff_f} and \eqref{eq:coeff_g}).

\section{Momentum diffusion constant} \label{sec:momdiff}
\subsection{Force-force correlator}
In this formulation, heavy quark is a quantum mechanical particle while the critical fluctuations are classical fields.
One should regard the classical critical fields as classical approximation of the corresponding quantum fields of critical fluctuations.
Then, the effective Lagrangian for the non-relativistic heavy quark with mass $M$ reads
\begin{align}
\mathcal L_{\rm Q} &= Q^{\dagger}\left(i\partial_t -M + \frac{\bm \nabla^2}{2M} - \lambda \delta \sigma\right)Q \nonumber \\
&\quad +Q_c^{\dagger}\left(i\partial_t -M + \frac{\bm \nabla^2}{2M} - \lambda_c\delta\sigma \right)Q_c.
\end{align}
From the heavy quark momentum operator
\begin{align}
\hat{\bm P}_{\rm Q} = \int d^3 x \left[\hat Q^{\dagger}\left(\frac{1}{i}\bm \nabla\right) \hat Q
 + \hat Q_c^{\dagger}\left(\frac{1}{i}\bm \nabla\right) \hat Q_c\right],
\end{align}
we can get the force operator
\begin{align}
\hat{\bm F}_{\rm Q} = \frac{d}{dt}\hat{\bm P}_{\rm Q} 
= -\int d^3 x \bm\nabla\delta\hat \sigma \cdot \left(\lambda \hat Q^{\dagger} \hat Q + \lambda_c \hat Q_c^{\dagger} \hat Q_c\right).
\end{align}
The heavy quark momentum diffusion constant $\kappa$ is defined by the force-force correlator \cite{Casalderrey-Solana:2006fio, Caron-Huot:2007rwy, Caron-Huot:2008dyw}
\begin{align}
\kappa &= \frac{1}{3} \int_{-\infty}^{\infty} dt \left\langle \hat{\bm F}_{\rm Q}(t) \cdot \hat{\bm F}_{\rm Q}(0) \right\rangle_{\rm Q},
\label{eq:kappa}
\end{align}
in the static limit of heavy quark ($M\to\infty$).
Here, $\langle\hat {\mathcal O}\rangle_{\rm Q}$ denotes a thermal average including one heavy quark put at $\bm x = \bm x_0$ in the infinite past:
\begin{align}
\langle\hat {\mathcal O}\rangle_{\rm Q} 
= \frac{\sum_n e^{-\beta(E_n - \mu B_n)}\langle n| \hat Q(\bm x_0, -\infty) \hat {\mathcal O} \hat Q^{\dagger}(\bm x_0, -\infty)|n\rangle}
{\sum_n e^{-\beta(E_n - \mu B_n)}},
\end{align}
where $E_n$ and $B_n$ are energy and baryon number of a QCD eigenstate $|n\rangle$ including no heavy quarks.
Drag force $-\gamma \bm p$ and diffusion constant $D$ are related to $\kappa$ by
\begin{align}
\gamma = \frac{\kappa}{2MT}, \quad D = \frac{2T^2}{\kappa}.
\end{align}

Apart from the rapid phase $e^{-iMt}$, which is absorbed by shifting the energy by $M$, the heavy quark field operator is easily solved in the static limit
\begin{align}
\hat Q(\bm x, t) &= \hat U(t,0;\bm x)\hat Q(\bm x, 0), \\
\hat U(t_1,t_2;\bm x) &\equiv {\rm P}e^{-i\int_{t_2}^{t_1} dt\lambda\delta\hat \sigma(\bm x, t)},
\end{align}
where $\rm P$ denotes the path-ordered product (i.e., when $t_1>t_2$ it is time-ordered product and when $t_1<t_2$ it is anti-time-ordered product).
Using the anti-commutator $\{\hat Q(\bm x, t) , \hat Q^{\dagger}(\bm y, t) \}=\delta(\bm x-\bm y)$ and the property $\hat Q(\bm x, t)|n\rangle = 0$, the force-force correlator in \eqref{eq:kappa} is expressed only with $\delta\hat\sigma$:
\begin{align}
\label{eq:FF}
&\left\langle \hat {\bm F}_{\rm Q}(t) \cdot \hat {\bm F}_{\rm Q}(0) \right\rangle_{\rm Q}\\
&=\lambda^2\left\langle
\hat U(-\infty,t)\bm\nabla\delta\hat\sigma(t)\hat U(t,0)\cdot \bm\nabla\delta\hat\sigma(0)\hat U(0,-\infty)
\right\rangle,  \nonumber 
\end{align} 
where $\langle\hat {\mathcal O}\rangle$ denotes thermal average without any heavy quarks.
Here, spatial position is dropped from all the operators because they are all at $\bm x_0$.

Near the critical point, the critical mode with long wavelength is responsible for the critical behaviors.
Occupation number of bosonic fields with low energy far exceeds 1 so that classical field approximation works excellently.
In the classical limit, $\delta\hat\sigma$ is no longer an operator and the phases of the transporters $U$s cancel out.
Thus, we obtain
\begin{align}
\left\langle \hat {\bm F}_{\rm Q}(t) \cdot \hat {\bm F}_{\rm Q}(0) \right\rangle_{\rm Q}
\simeq \lambda^2\left\langle
\bm\nabla\delta\sigma(t)\cdot \bm\nabla\delta\sigma(0)
\right\rangle. 
\end{align} 
Two point function of the classical field $\delta\sigma$ can be calculated using the generalized Langevin equation for the model H \cite{RevModPhys.49.435}.

Here are two technical remarks on the classical approximation.
First, to be strict, it is necessary to include the back reaction from the classical critical field to the heavy quark.
The transporters $\hat U$ can be included in the Hamiltonian as $\lambda \int d^3x \delta(\bm x-\bm x_0)\delta\hat\sigma(\bm x)$ so that the classical equation of motion has a new source term at the position of the heavy quark.
In this sense, our result would be strictly justified when the coupling $\lambda$ is weak.
Second, the amplitude of the classical field $\delta\sigma$ becomes smaller near the critical point.
Up to a small scaling exponent $\eta\sim 0.04$, one can estimate it from the susceptibility
\begin{align}
\int d^3xd^3y \langle\delta\sigma(\bm x)\delta\sigma(\bm y)\rangle &\sim  VT\xi^{2}  
\sim \frac{V}{\xi^3}\cdot (\xi^3)^2 \cdot \frac{T}{\xi},
\end{align}
where $V$ is the system volume and $\xi$ is the correlation length.
On the right hand side, $V/\xi^3$ is a number of independent domains, $(\xi^3)^2$ is from the volume measure $d^3xd^3y$, and thus $\sqrt{T/\xi}$ is the typical amplitude of $\delta\sigma$ with wavelength $\xi$.
Comparing the energy density $(\bm\nabla\delta\sigma)^2 \sim T/\xi^3$ and the interaction energy by the source $\lambda\delta\sigma(\bm x_0)\sim \lambda \sqrt{T/\xi}$, the effect of source should be localized within a volume  $v \lesssim \lambda T^{-1/2}\xi^{5/2} \ll \xi^3$.
This observation may support that the back reaction can be generally negligible near the critical point, but it deserves further analysis.

\subsection{Scaling behavior}
Scaling behavior of (connected) correlation function
\begin{align}
C(\bm x,t) &\equiv \langle\delta\sigma(\bm x,t)\delta\sigma(\bm 0,0)\rangle_\text{conn}\nonumber \\
&=\int\frac{d^3q}{(2\pi)^3} e^{i\bm q\cdot\bm x}\tilde C(\bm q,t)
\end{align}
is well known.
It is related to the susceptibility $\chi(q)$ and the relaxation time $\tau(q)$ by
\begin{align}
&\tilde C(\bm q, 0) = \chi(q), \quad
\int dt \tilde C(\bm q, t) \sim \chi(q)\tau(q).
\end{align}
Therefore, we get
\begin{align}
\kappa &= \frac{\lambda^2}{3} \int dt \int \frac{d^3q}{(2\pi)^3}  \tilde C(\bm q,t)q^2
\sim \frac{\lambda^2}{3} \int d^3q \chi(q)\tau(q) q^2.
\end{align}
The susceptibility $\chi(q)$ and relaxation time $\tau(q)$ obey the scaling form
\begin{align}
\chi(q) \sim q^{-2+\eta}, \quad \tau(q) \sim q^{-z},
\end{align}
in the scaling regime $1/\xi \ll q \ll 1/r_o$, where $\xi$ is the correlation length and $r_o$ is some microscopic cutoff length scale.
Thus, the integral becomes
\begin{align}
\kappa \sim  \frac{\lambda^2}{3} \int_{1/\xi}^{1/r_o} dq q^{2+\eta-z}
\sim  \frac{\lambda^2}{3}\left(\frac{r_o^{z-3-\eta} - \xi^{z-3-\eta}}{3+\eta-z}\right).
\end{align}
When $z-3-\eta > 0$, the integral is dominated by the fluctuations with long wavelength and hence $\kappa\propto \xi^{z-3-\eta}$ is sensitive to the correlation length.
When $z-3-\eta < 0$, the integral is dominated by the fluctuations with short wavelength and hence $\kappa\propto r_o^{-|z-3-\eta|}$ does not exhibit critical scaling.
In the latter case, the critical fluctuation does not make a dominant contribution to $\kappa$.
Rather, $\kappa$ is determined by microscopic collisional process, which is insensitive to the critical behavior.

It is predicted that the critical dynamics of the QCD critical point is given by the model H in the Hohenberg-Halperin classification \cite{Son:2004iv, RevModPhys.49.435}, for which $z\simeq 3$ and $\eta\sim 0.04$.
In this case, we conclude that the critical behavior is not significant and is difficult to find in the heavy-ion collisions.
In the case of model B, $z\simeq 4$ and $\kappa\sim \xi$ shows singular behavior near the critical point, but the model B scenario is somewhat unlikely.

\section{Conclusion} \label{sec:conclusion}

As we have seen, in the case of model H the critical exponent is unexpectedly small and it is difficult to observe critical phenomenon through $v_2$ of heavy quarks.
However, it is interesting that in the case of model B it could be observable.
The difference between model B and model H is only one additional conserved quantity (in our case momentum density) and existence of Poissonian dynamics.
This difference makes the observability of heavy quark $v_2$ thus different.
It would be meaningful to confirm this difference in atomic systems in laboratory.
In addition, recently the effect of the chiral critical point at $m=0$ has been studied on the thermodynamical behavior at the physical point and observed on the lattice \cite{HotQCD:2019xnw}.
This critical point has $\text{O}(4)$ symmetry and belongs to model G \cite{Rajagopal:1992qz, Florio:2021jlx, Florio:2023kmy}. 
Thus, it would be of interest to extend our analysis to model G.
Finally, there may be a possibility that heavy quark strongly couples with light antiquarks near the critical point and forms a composite particle, which is no longer singlet in the chiral symmetry.
In this case, construction of the effective interaction Lagrangian would be more involved and the force-force correlation function remains complicated even in the classical limit because of the non-Abelian nature of chiral symmetry.

\section*{Acknowledgements} \label{sec:acknowledgements}
Y.A. thanks Masaru Hongo for fruitful discussions.
Y.A. is supported by JSPS KAKENHI Grant Numbers JP18K13538 and JP23H01174.
M.A. is supported in part by JSPS KAKENHI Grant Numbers JP21H00124 and JP23K03386.

\bibliography{Refs}

\begin{thebibliography}{15}%
\makeatletter
\providecommand \@ifxundefined [1]{%
 \@ifx{#1\undefined}
}%
\providecommand \@ifnum [1]{%
 \ifnum #1\expandafter \@firstoftwo
 \else \expandafter \@secondoftwo
 \fi
}%
\providecommand \@ifx [1]{%
 \ifx #1\expandafter \@firstoftwo
 \else \expandafter \@secondoftwo
 \fi
}%
\providecommand \natexlab [1]{#1}%
\providecommand \enquote  [1]{``#1''}%
\providecommand \bibnamefont  [1]{#1}%
\providecommand \bibfnamefont [1]{#1}%
\providecommand \citenamefont [1]{#1}%
\providecommand \href@noop [0]{\@secondoftwo}%
\providecommand \href [0]{\begingroup \@sanitize@url \@href}%
\providecommand \@href[1]{\@@startlink{#1}\@@href}%
\providecommand \@@href[1]{\endgroup#1\@@endlink}%
\providecommand \@sanitize@url [0]{\catcode `\\12\catcode `\$12\catcode
  `\&12\catcode `\#12\catcode `\^12\catcode `\_12\catcode `\%12\relax}%
\providecommand \@@startlink[1]{}%
\providecommand \@@endlink[0]{}%
\providecommand \url  [0]{\begingroup\@sanitize@url \@url }%
\providecommand \@url [1]{\endgroup\@href {#1}{\urlprefix }}%
\providecommand \urlprefix  [0]{URL }%
\providecommand \Eprint [0]{\href }%
\providecommand \doibase [0]{http://dx.doi.org/}%
\providecommand \selectlanguage [0]{\@gobble}%
\providecommand \bibinfo  [0]{\@secondoftwo}%
\providecommand \bibfield  [0]{\@secondoftwo}%
\providecommand \translation [1]{[#1]}%
\providecommand \BibitemOpen [0]{}%
\providecommand \bibitemStop [0]{}%
\providecommand \bibitemNoStop [0]{.\EOS\space}%
\providecommand \EOS [0]{\spacefactor3000\relax}%
\providecommand \BibitemShut  [1]{\csname bibitem#1\endcsname}%
\let\auto@bib@innerbib\@empty
\bibitem [{\citenamefont {Arslandok}\ \emph {et~al.}(2023)\citenamefont
  {Arslandok} \emph {et~al.}}]{Arslandok:2023utm}%
  \BibitemOpen
  \bibfield  {author} {\bibinfo {author} {\bibfnamefont {M.}~\bibnamefont
  {Arslandok}} \emph {et~al.},\ }\href@noop {} {\  (\bibinfo {year} {2023})},\
  \Eprint {http://arxiv.org/abs/2303.17254} {arXiv:2303.17254 [nucl-ex]}
  \BibitemShut {NoStop}%
\bibitem [{\citenamefont {Asakawa}\ and\ \citenamefont
  {Yazaki}(1989)}]{Asakawa:1989bq}%
  \BibitemOpen
  \bibfield  {author} {\bibinfo {author} {\bibfnamefont {M.}~\bibnamefont
  {Asakawa}}\ and\ \bibinfo {author} {\bibfnamefont {K.}~\bibnamefont
  {Yazaki}},\ }\href {\doibase 10.1016/0375-9474(89)90002-X} {\bibfield
  {journal} {\bibinfo  {journal} {Nucl. Phys. A}\ }\textbf {\bibinfo {volume}
  {504}},\ \bibinfo {pages} {668} (\bibinfo {year} {1989})}\BibitemShut
  {NoStop}%
\bibitem [{\citenamefont {Asakawa}\ \emph {et~al.}(2020)\citenamefont
  {Asakawa}, \citenamefont {Kitazawa},\ and\ \citenamefont
  {M\"uller}}]{Asakawa:2019kek}%
  \BibitemOpen
  \bibfield  {author} {\bibinfo {author} {\bibfnamefont {M.}~\bibnamefont
  {Asakawa}}, \bibinfo {author} {\bibfnamefont {M.}~\bibnamefont {Kitazawa}}, \
  and\ \bibinfo {author} {\bibfnamefont {B.}~\bibnamefont {M\"uller}},\ }\href
  {\doibase 10.1103/PhysRevC.101.034913} {\bibfield  {journal} {\bibinfo
  {journal} {Phys. Rev. C}\ }\textbf {\bibinfo {volume} {101}},\ \bibinfo
  {pages} {034913} (\bibinfo {year} {2020})},\ \Eprint
  {http://arxiv.org/abs/1912.05840} {arXiv:1912.05840 [nucl-th]} \BibitemShut
  {NoStop}%
\bibitem [{\citenamefont {Hohenberg}\ and\ \citenamefont
  {Halperin}(1977)}]{RevModPhys.49.435}%
  \BibitemOpen
  \bibfield  {author} {\bibinfo {author} {\bibfnamefont {P.~C.}\ \bibnamefont
  {Hohenberg}}\ and\ \bibinfo {author} {\bibfnamefont {B.~I.}\ \bibnamefont
  {Halperin}},\ }\href {\doibase 10.1103/RevModPhys.49.435} {\bibfield
  {journal} {\bibinfo  {journal} {Rev. Mod. Phys.}\ }\textbf {\bibinfo {volume}
  {49}},\ \bibinfo {pages} {435} (\bibinfo {year} {1977})}\BibitemShut
  {NoStop}%
\bibitem [{\citenamefont {Son}\ and\ \citenamefont
  {Stephanov}(2004)}]{Son:2004iv}%
  \BibitemOpen
  \bibfield  {author} {\bibinfo {author} {\bibfnamefont {D.~T.}\ \bibnamefont
  {Son}}\ and\ \bibinfo {author} {\bibfnamefont {M.~A.}\ \bibnamefont
  {Stephanov}},\ }\href {\doibase 10.1103/PhysRevD.70.056001} {\bibfield
  {journal} {\bibinfo  {journal} {Phys. Rev. D}\ }\textbf {\bibinfo {volume}
  {70}},\ \bibinfo {pages} {056001} (\bibinfo {year} {2004})},\ \Eprint
  {http://arxiv.org/abs/hep-ph/0401052} {arXiv:hep-ph/0401052} \BibitemShut
  {NoStop}%
\bibitem [{\citenamefont {Brown}(1828)}]{brown1828xxvii}%
  \BibitemOpen
  \bibfield  {author} {\bibinfo {author} {\bibfnamefont {R.}~\bibnamefont
  {Brown}},\ }\href@noop {} {\bibfield  {journal} {\bibinfo  {journal} {The
  philosophical magazine}\ }\textbf {\bibinfo {volume} {4}},\ \bibinfo {pages}
  {161} (\bibinfo {year} {1828})}\BibitemShut {NoStop}%
\bibitem [{\citenamefont {Rajagopal}\ and\ \citenamefont
  {Wilczek}(1993)}]{Rajagopal:1992qz}%
  \BibitemOpen
  \bibfield  {author} {\bibinfo {author} {\bibfnamefont {K.}~\bibnamefont
  {Rajagopal}}\ and\ \bibinfo {author} {\bibfnamefont {F.}~\bibnamefont
  {Wilczek}},\ }\href {\doibase 10.1016/0550-3213(93)90502-G} {\bibfield
  {journal} {\bibinfo  {journal} {Nucl. Phys. B}\ }\textbf {\bibinfo {volume}
  {399}},\ \bibinfo {pages} {395} (\bibinfo {year} {1993})},\ \Eprint
  {http://arxiv.org/abs/hep-ph/9210253} {arXiv:hep-ph/9210253} \BibitemShut
  {NoStop}%
\bibitem [{\citenamefont {Fujii}\ and\ \citenamefont
  {Ohtani}(2004)}]{Fujii:2004jt}%
  \BibitemOpen
  \bibfield  {author} {\bibinfo {author} {\bibfnamefont {H.}~\bibnamefont
  {Fujii}}\ and\ \bibinfo {author} {\bibfnamefont {M.}~\bibnamefont {Ohtani}},\
  }\href {\doibase 10.1103/PhysRevD.70.014016} {\bibfield  {journal} {\bibinfo
  {journal} {Phys. Rev. D}\ }\textbf {\bibinfo {volume} {70}},\ \bibinfo
  {pages} {014016} (\bibinfo {year} {2004})},\ \Eprint
  {http://arxiv.org/abs/hep-ph/0402263} {arXiv:hep-ph/0402263} \BibitemShut
  {NoStop}%
\bibitem [{\citenamefont {Chin}(1977)}]{CHIN1977301}%
  \BibitemOpen
  \bibfield  {author} {\bibinfo {author} {\bibfnamefont {S.}~\bibnamefont
  {Chin}},\ }\href {\doibase https://doi.org/10.1016/0003-4916(77)90016-1}
  {\bibfield  {journal} {\bibinfo  {journal} {Annals of Physics}\ }\textbf
  {\bibinfo {volume} {108}},\ \bibinfo {pages} {301} (\bibinfo {year}
  {1977})}\BibitemShut {NoStop}%
\bibitem [{\citenamefont {Casalderrey-Solana}\ and\ \citenamefont
  {Teaney}(2006)}]{Casalderrey-Solana:2006fio}%
  \BibitemOpen
  \bibfield  {author} {\bibinfo {author} {\bibfnamefont {J.}~\bibnamefont
  {Casalderrey-Solana}}\ and\ \bibinfo {author} {\bibfnamefont
  {D.}~\bibnamefont {Teaney}},\ }\href {\doibase 10.1103/PhysRevD.74.085012}
  {\bibfield  {journal} {\bibinfo  {journal} {Phys. Rev. D}\ }\textbf {\bibinfo
  {volume} {74}},\ \bibinfo {pages} {085012} (\bibinfo {year} {2006})},\
  \Eprint {http://arxiv.org/abs/hep-ph/0605199} {arXiv:hep-ph/0605199}
  \BibitemShut {NoStop}%
\bibitem [{\citenamefont {Caron-Huot}\ and\ \citenamefont
  {Moore}(2008{\natexlab{a}})}]{Caron-Huot:2007rwy}%
  \BibitemOpen
  \bibfield  {author} {\bibinfo {author} {\bibfnamefont {S.}~\bibnamefont
  {Caron-Huot}}\ and\ \bibinfo {author} {\bibfnamefont {G.~D.}\ \bibnamefont
  {Moore}},\ }\href {\doibase 10.1103/PhysRevLett.100.052301} {\bibfield
  {journal} {\bibinfo  {journal} {Phys. Rev. Lett.}\ }\textbf {\bibinfo
  {volume} {100}},\ \bibinfo {pages} {052301} (\bibinfo {year}
  {2008}{\natexlab{a}})},\ \Eprint {http://arxiv.org/abs/0708.4232}
  {arXiv:0708.4232 [hep-ph]} \BibitemShut {NoStop}%
\bibitem [{\citenamefont {Caron-Huot}\ and\ \citenamefont
  {Moore}(2008{\natexlab{b}})}]{Caron-Huot:2008dyw}%
  \BibitemOpen
  \bibfield  {author} {\bibinfo {author} {\bibfnamefont {S.}~\bibnamefont
  {Caron-Huot}}\ and\ \bibinfo {author} {\bibfnamefont {G.~D.}\ \bibnamefont
  {Moore}},\ }\href {\doibase 10.1088/1126-6708/2008/02/081} {\bibfield
  {journal} {\bibinfo  {journal} {JHEP}\ }\textbf {\bibinfo {volume} {02}},\
  \bibinfo {pages} {081} (\bibinfo {year} {2008}{\natexlab{b}})},\ \Eprint
  {http://arxiv.org/abs/0801.2173} {arXiv:0801.2173 [hep-ph]} \BibitemShut
  {NoStop}%
\bibitem [{\citenamefont {Ding}\ \emph {et~al.}(2019)\citenamefont {Ding} \emph
  {et~al.}}]{HotQCD:2019xnw}%
  \BibitemOpen
  \bibfield  {author} {\bibinfo {author} {\bibfnamefont {H.~T.}\ \bibnamefont
  {Ding}} \emph {et~al.} (\bibinfo {collaboration} {HotQCD}),\ }\href {\doibase
  10.1103/PhysRevLett.123.062002} {\bibfield  {journal} {\bibinfo  {journal}
  {Phys. Rev. Lett.}\ }\textbf {\bibinfo {volume} {123}},\ \bibinfo {pages}
  {062002} (\bibinfo {year} {2019})},\ \Eprint
  {http://arxiv.org/abs/1903.04801} {arXiv:1903.04801 [hep-lat]} \BibitemShut
  {NoStop}%
\bibitem [{\citenamefont {Florio}\ \emph {et~al.}(2022)\citenamefont {Florio},
  \citenamefont {Grossi}, \citenamefont {Soloviev},\ and\ \citenamefont
  {Teaney}}]{Florio:2021jlx}%
  \BibitemOpen
  \bibfield  {author} {\bibinfo {author} {\bibfnamefont {A.}~\bibnamefont
  {Florio}}, \bibinfo {author} {\bibfnamefont {E.}~\bibnamefont {Grossi}},
  \bibinfo {author} {\bibfnamefont {A.}~\bibnamefont {Soloviev}}, \ and\
  \bibinfo {author} {\bibfnamefont {D.}~\bibnamefont {Teaney}},\ }\href
  {\doibase 10.1103/PhysRevD.105.054512} {\bibfield  {journal} {\bibinfo
  {journal} {Phys. Rev. D}\ }\textbf {\bibinfo {volume} {105}},\ \bibinfo
  {pages} {054512} (\bibinfo {year} {2022})},\ \Eprint
  {http://arxiv.org/abs/2111.03640} {arXiv:2111.03640 [hep-lat]} \BibitemShut
  {NoStop}%
\bibitem [{\citenamefont {Florio}\ \emph {et~al.}(2023)\citenamefont {Florio},
  \citenamefont {Grossi},\ and\ \citenamefont {Teaney}}]{Florio:2023kmy}%
  \BibitemOpen
  \bibfield  {author} {\bibinfo {author} {\bibfnamefont {A.}~\bibnamefont
  {Florio}}, \bibinfo {author} {\bibfnamefont {E.}~\bibnamefont {Grossi}}, \
  and\ \bibinfo {author} {\bibfnamefont {D.}~\bibnamefont {Teaney}},\
  }\href@noop {} {\  (\bibinfo {year} {2023})},\ \Eprint
  {http://arxiv.org/abs/2306.06887} {arXiv:2306.06887 [hep-lat]} \BibitemShut
  {NoStop}%
\end{thebibliography}%


\end{document}